\documentclass[submission,copyright,creativecommons]{eptcs}

\providecommand{\event}{TFPIE 2016} 
\usepackage{breakurl}             

\usepackage[english]{babel}
\usepackage{color}
\usepackage{graphicx}

\usepackage{listings}
\usepackage{textcomp}

\usepackage{fancybox}
\usepackage{fancyhdr}

\usepackage{caption}
\usepackage{subcaption}

\usepackage{booktabs}

\usepackage{lipsum}
\usepackage{verbatim}

\newcommand{\bricklayer}{\emph{Bricklayer}}
\newcommand{\lego}{LEGO\ensuremath{^{\textnormal{\textregistered}}}}

\definecolor{keywordColor}{rgb}{0.5,0.00,0.25}
\definecolor{commentColor}{rgb}{0.00,0.50,0.00}
\definecolor{stringColor}{rgb}{0.00,0.00,1.00}
\definecolor{bgColor}{rgb}{0.85,0.95,1.00}

\def\titlerunning{The \bricklayer\ Ecosystem}
\def\authorrunning{Winter - Love - Corritore}

\begin{document}

\title{The Bricklayer Ecosystem - Art, Math, and Code}

\author{Victor Winter
\institute{Department of Computer Science\\
University of Nebraska at Omaha\\
Omaha, NE 68182, USA}
\email{vwinter@unomaha.edu}
\and{Betty Love}
\institute{Department of Mathematics\\
University of Nebraska at Omaha\\
Omaha, NE 68182, USA}
\email{blove@unomaha.edu}
\and{Cindy Corritore}
\institute{Creighton University \\
Omaha, NE 68102, USA}
\email{cindycc@gmail.com}
}

\maketitle
\begin{abstract}
This paper describes the \bricklayer\ Ecosystem -- a freely-available online educational ecosystem created for people of all ages and coding backgrounds. \bricklayer\ is designed in accordance with a ``low-threshold infinite ceiling'' philosophy and has been successfully used to teach coding to primary school students, middle school students, university freshmen, and in-service secondary math teachers. \bricklayer\ programs are written in the functional programming language SML and, when executed, create 2D and 3D artifacts. These artifacts can be viewed using a variety of third-party tools such as \lego\ Digital Designer (LDD), LDraw, Minecraft clients, Brickr, as well as STereoLithography viewers.
\end{abstract}





\section{Motivation}

Global interest in teaching coding across the entire K-12 spectrum is rapidly increasing. A number of countries in the EU, including England \cite{dredge-england}, currently teach computational thinking and coding starting in kindergarten. A noteworthy recent addition (2016) to this list is Finland, whose educational system is consistently ranked among the highest in the Western world and has been touted as taking coding education farther than any country in Europe \cite{toikkanen-finland}. On the other side of the globe, Australia now also teaches coding across the K-12 spectrum. Australia is interesting because it has made time for coding in its curriculum by eliminating history and geography \cite{chang-aussie}.

The educational landscape in the United States is less unified. However, progress is being made. Cities such as San Francisco, New York, and Chicago have plans to teach coding across the K-12 spectrum. Recently, Rhode Island has rolled out a comprehensive effort, called CS4RI, to teach coding across the entire K-12 spectrum statewide \cite{raimondo-rhode-island}.

The pursuit of coding in the US is also being strongly encouraged at the national level. In his 2016 State of the Union Address, President Obama presented his \emph{Computer Science for All} initiative that calls for \$4 billion in funding for K-12 computer science education~\cite{smith2016}. This interest in computer science is reflected in similar initiatives across the United States and around the world~\cite{pretz2014,gross2015,wheatley2016}. The purpose is not only to provide students with the programming skills needed for high-tech careers, but more generally to provide them with the computational thinking skills they will need for careers in a technology-driven world.

The resulting educational environment has given rise to an increased exploration and development of educational tools and systems whose purpose is to teach coding within the K-12 spectrum. The \bricklayer\ ecosystem is one such system.

The remainder of the paper is as follows: Section \ref{section-related-work} gives an overview of some existing educational tools and systems. Section \ref{section-bricklayer-ecosystem} describes how \bricklayer\ has evolved since its debut at TFPIE 2014. Section \ref{section-additions} describes the components that make up the \bricklayer\ ecosystem. Section \ref{section-education} describes some of the settings in which \bricklayer\ is being used, and Section \ref{section-conclusion} concludes.

\section{Related Work}\label{section-related-work}

This section describes 3 major educational systems targeting K12 education. Each system offers a distinct perspective on how to effectively engage K-12 students in coding. We omit discussion of systems such as Alice, Snap!, Greenfoot, Media Computation, Picturing Programs, Yampa, and Soccer-fun all of which target audiences outside the K-12 range (e.g., high school and above).

\textbf{Scratch} \cite{Maloney:2010:Scratch} is a visual programming language and environment designed to teach programming to young people (8-16) especially those who have no previous programming experience. Building on the ideas of Logo, Scratch seeks to create a constructionist learning environment where young people can pursue projects that are both personalized and meaningful. Learning is achieved through experimentation and peer sharing rather than traditional instruction. To accomplish these goals, the Scratch language has been designed so that it is virtually impossible to create an ill-formed program. They describe the rationale for this design decision in terms of the following \lego\ metaphor: \emph{``The [\lego] brick shapes suggest what is possible, and experimentation and experience teaches what works''}\cite{Maloney:2010:Scratch}. On the syntactic level language constructs are colorful blocks whose shape govern their composition. On the semantic level, Scratch employs data coercion and pre-defined ranges for data values to virtually eliminate all runtime errors. The result is a programming environment that encourages \emph{tinkering} with programs. This mindset is further enhanced by Scratch's \emph{liveness} property which enables users to observe the functionality of a block or program fragment by simply clicking on it and to modify programs during runtime.

\textbf{Bootstrap} \cite{Schanzer:2015:Bootstrap} is a programming language, environment and curriculum designed primarily for middle and high school students. Programs are written in a subset of Racket, a functional programming language, and focus on the creation of video games involving four principle abstractions: (1) a player, (2) a target, (3) a danger, and (4) a background. The video games created are deconstructed into a sequence of frames that center around a player trying to acquire targets while avoiding dangers -- all of which can change from one frame to the next. Programming involves the creation and modification of functions that define the rules of the game and govern the behavior of the player, targets, and dangers. A platform called the \emph{World framework} \cite{Felleisen:2009:FIS:1596550.1596561} is responsible for creating frame sequences (i.e., a video game) from a given set of functions. In this way, programmers are shielded from having to use looping constructs.

Bootstrap coding is designed to achieve very precise learning objectives, namely, to improve abilities in solving algebra word problems of the kind found in standardized algebra tests given to middle school students in the US. The Bootstrap philosophy is based on the premise that in order for transfer of learning to occur between domains (e.g., coding and algebra) study must be carefully guided. Towards this end, students are taught how to use a process called the \emph{(Bootstrap) Design Recipe} \cite{Felleisen:How-to-Design-Programs} to design and develop the functions that underly their video games. This approach has strong ties to key concepts defined in algebra learning standards.

It is worth mentioning that the abstractions of functional programming languages (e.g., variables and functions) are much more closely aligned with their mathematical counterparts than their imperative language counterparts. Thus, along this dimension, functional languages are better positioned for transfer of learning between the coding and algebra domains.

\textbf{CodeSpells} \cite{Esper:2013:CodeSpells}\cite{Esper:2013:NFS:2445196.2445290} is an extensible educational platform designed to teach Java programming in the context of video game play (a JavaScript/Blockly version was recently made available on Steam). CodeSpells targets learning spaces outside of the traditional classroom. The authors believe that it is in such non-institutionalized settings that lifelong passion for coding is most likely to develop.

In contrast to educational systems like Bootstrap, which focuses on \emph{building} video games, the focus of CodeSpells is on \emph{playing} a video game. More specifically, CodeSpells is a non-competitive role playing game in which the student is a wizard in a 3D world populated by gnome-like creatures. The game provides 7 primary spells which include levitation of objects, fire, and flying. Programming consists of modifying and creating spells, which are written in Java. Game play is largely unstructured, though quests can be undertaken to earn badges. The first level of the game introduces students to Java basics like method parameters, conditional statements, and iteration.

Codespells is suitable for both primary and secondary school students. The authors report that the immersive and role playing aspects of CodeSpells has a number of beneficial effects. Foremost among these are (1) countless hours -- video games have an ability to engage players for hours on end, and (2) positive attitude toward failure -- within the context of game play (a non-institutional setting) when a student creates code containing an error they do not see themselves as having done something \emph{wrong} -- rather, such errors are seen as part of the game.

\section{The Bricklayer Ecosystem}\label{section-bricklayer-ecosystem}
At its core, \bricklayer\ is a library of graphical primitives, written in SML, which can be used to create pixelated artifacts. A \bricklayer\ program is a program, written in the functional programming language SML, which references elements (e.g., functions) belonging to the \bricklayer\ library.

\begin{figure}[htb!]
\centering
\begin{subfigure}[b]{0.4\textwidth}
\includegraphics[width=0.9\textwidth]{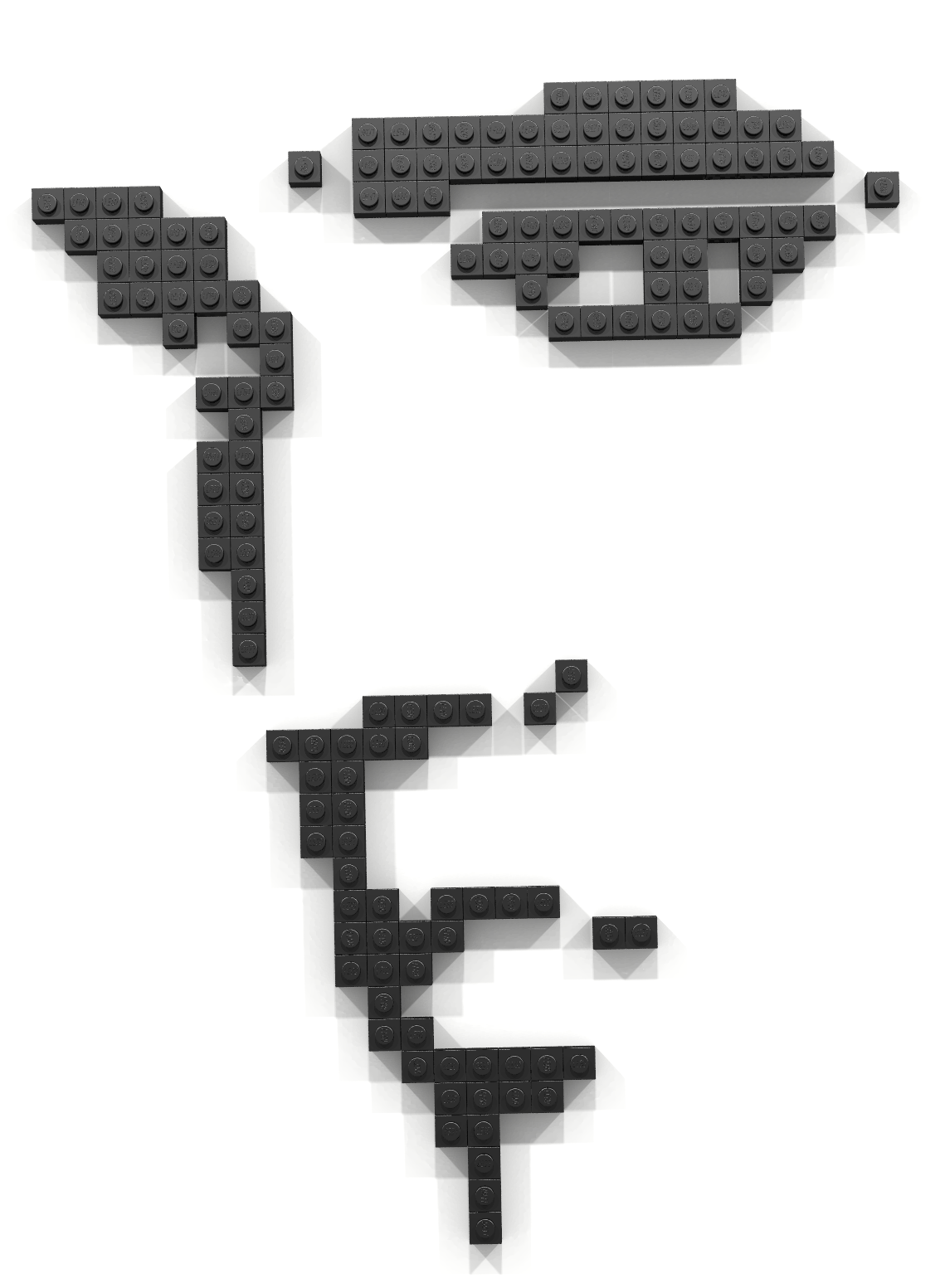}
\caption{A pixel art image of Elvis Presley.}
\label{fig:elvis}
\end{subfigure}
\hspace*{20mm}
\begin{subfigure}[b]{0.4\textwidth}
\includegraphics[width=1.0\textwidth]{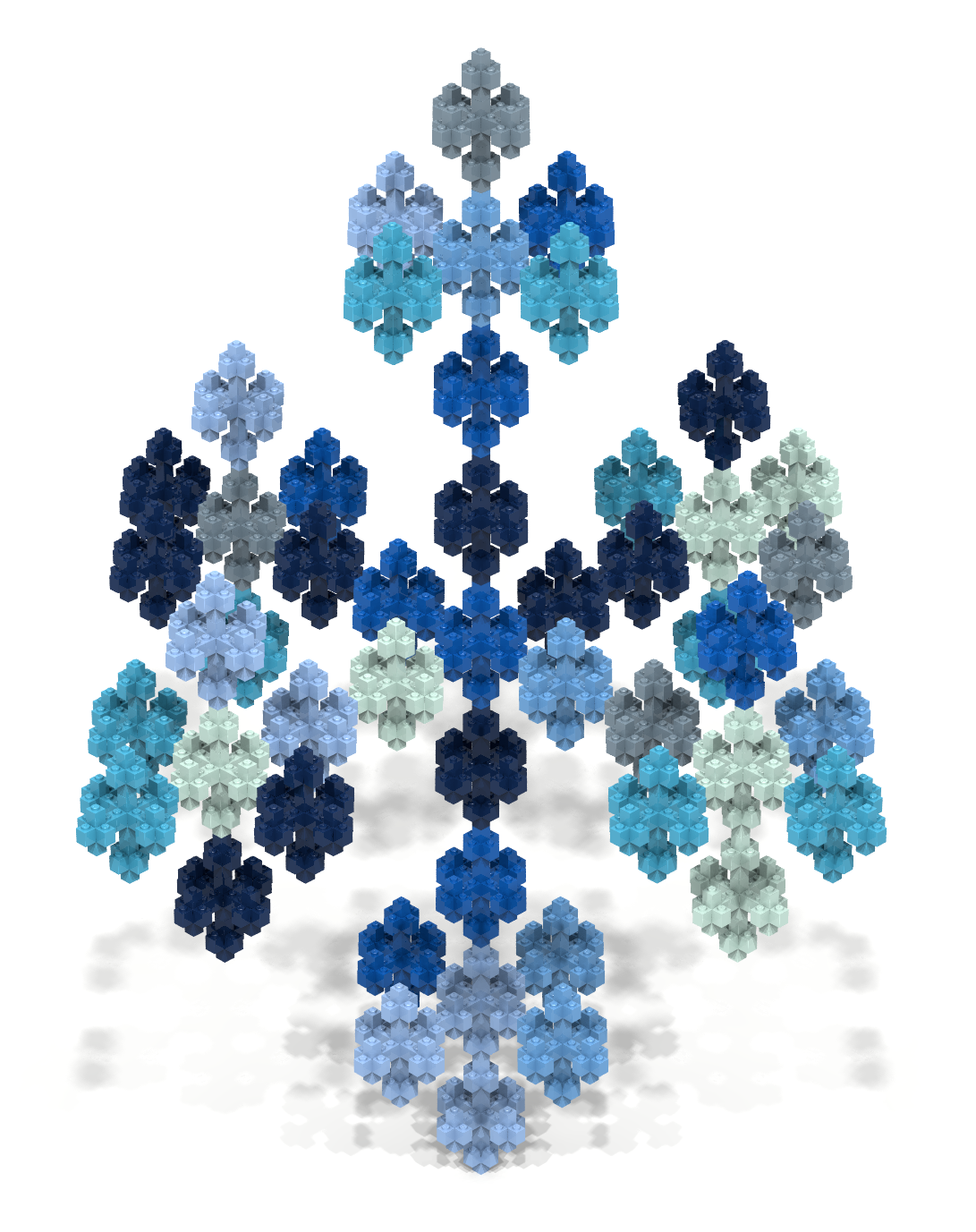}
\caption{The dual of the Menger sponge.}
\label{fig:dualMenger}
\end{subfigure}

\caption{Artifacts created by \bricklayer\ and rendered using POV-Ray.}~\label{fig:bricklayer-artifacts}
\end{figure}

\lego\ Digital Designer (LDD), a freely available tool provided by The \lego\ Group, is the default viewer for the pixelated artifacts created by \bricklayer\ programs. Figure \ref{fig:bricklayer-artifacts} shows \lego\ artifacts created by \bricklayer\ programs and rendered using the Persistence of Vision Raytracer (POV-Ray)\cite{povray} tool.

\subsection{Then}

The \bricklayer\ library was debuted at TFPIE 2014 \cite{2014Winter:TFPIE-EPTCS}. At that time, the library consisted of predicates, iterators, and around 30 \lego\ bricks. An abstraction called a \emph{brick function} was also provided which is similar to a predicate except that its return value is a \lego\ brick rather than a Boolean value. In 2014, the construction of \bricklayer\ artifacts was limited to compositions involving these primitives.

Though highly expressive from a theoretical standpoint, the sophistication of these abstractions presented a barrier to students in the K-12 spectrum who wanted to learn how to code. For example, it is simply too much to expect elementary school students to have fluency in Boolean algebra in order to create Bricklayer artifacts. To address such limitations, a major redesign of \bricklayer\ was undertaken with the goal of making \bricklayer\ more accessible across the K-12 spectrum. It is this redesign that is the focus of this paper.

\subsection{Now}

At present, the \bricklayer\ library consists of five coding levels (with two more levels in the design stage). Levels 1--3 provide primitives for creating two dimensional artifacts, and levels 4--5 provide primitives for creating three dimensional artifacts. The predicates, iterators, and brick functions that were available in the version of \bricklayer\ reported on in TFPIE 2014 now reside in level 5.

In addition, the new \bricklayer\ library is bundled with a collection of interactive web apps, online videos, documentation, coding examples, and exercises whose collective purpose is to help people of all ages and coding backgrounds learn how to write \bricklayer\ programs. We refer to this collection as the \bricklayer\ ecosystem -- an ecosystem which has been designed in accordance with a ``low-threshold infinite-ceiling'' philosophy. The \bricklayer\ ecosystem can be found at the following URL.

\begin{center}
\url{bricklayer.org}
\end{center}

\section{Additions to \bricklayer}\label{section-additions}

At TFPIE 2014 numerous suggestions were made regarding ways that \bricklayer\ might be extended or enhanced. In this section we give an overview of the extensions (some that were proposed, some not) that we have made to the original \bricklayer\ library as well as to the \bricklayer\ ecosystem.

\subsection{Vitruvia}\label{section-vitruvia}

\begin{figure}[htb!]
\centering
\begin{subfigure}[b]{0.8\textwidth}
\includegraphics[width=\textwidth]{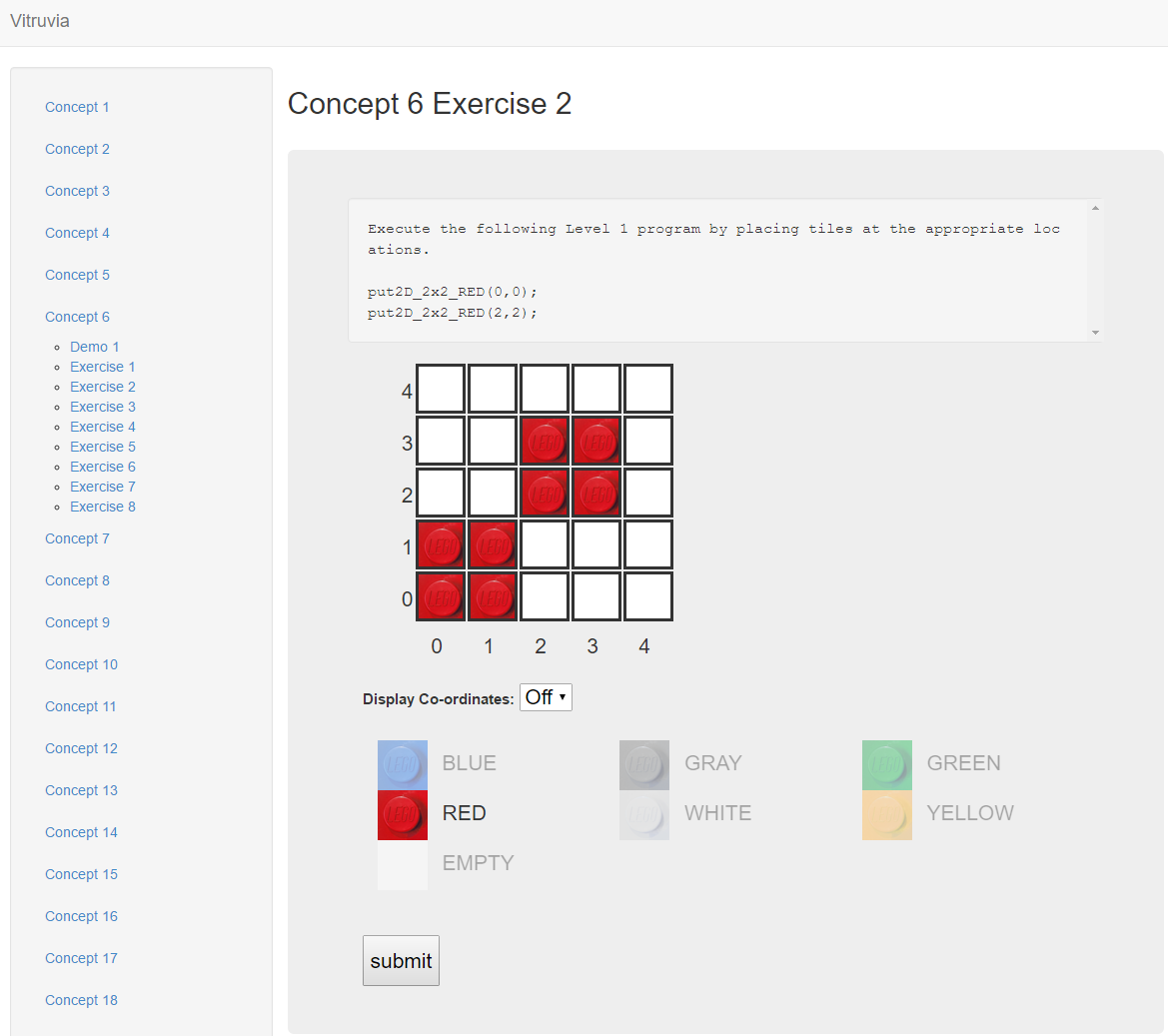}
\caption{Solution submitted by student.}
\label{fig:vitruvia-a}
\end{subfigure}

\bigskip

\begin{subfigure}[b]{0.8\textwidth}
\includegraphics[width=\textwidth]{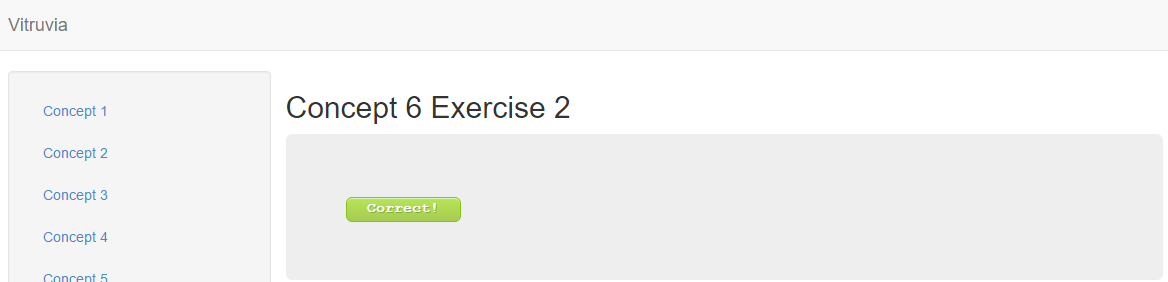}
\caption{Automated feedback provided by Vitruvia.}
\label{fig:vitruvia-b}
\end{subfigure}
\caption{An example of a Vitruvia exercise.}~\label{fig:vitruvia}
\end{figure}

Vitruvia\footnote{Legend has it that Marcus Vitruvis had a daughter named Vitruvia.} is an interactive web app that is part of the \bricklayer\ ecosystem. In Vitruvia, coding skills can be developed through exercises which require input from the user (via mouse clicks or touch screen selections) and which are automatically graded. These interactive exercises are grouped into \emph{concepts}. Figure \ref{fig:vitruvia} shows an example of a manually created solution to an exercise and the automated feedback provided by Vitruvia.

Presently, Vitruvia contains 23 concepts (more are in the design stage) which provide a gentle learning curve accessible to people of all ages and backgrounds. In our TFPIE 2014 paper we discussed characteristics of effective example-problem sequences and asserted that the graphical domain of \bricklayer, which we described as being \emph{example rich and problem dense}, was well suited to the creation of such sequences \cite{2014Winter:TFPIE-EPTCS}. The concept sequence in Vitruvia is an instance of such an example-problem sequence.

\begin{figure}[htb!]
\centering
\includegraphics[width=\textwidth]{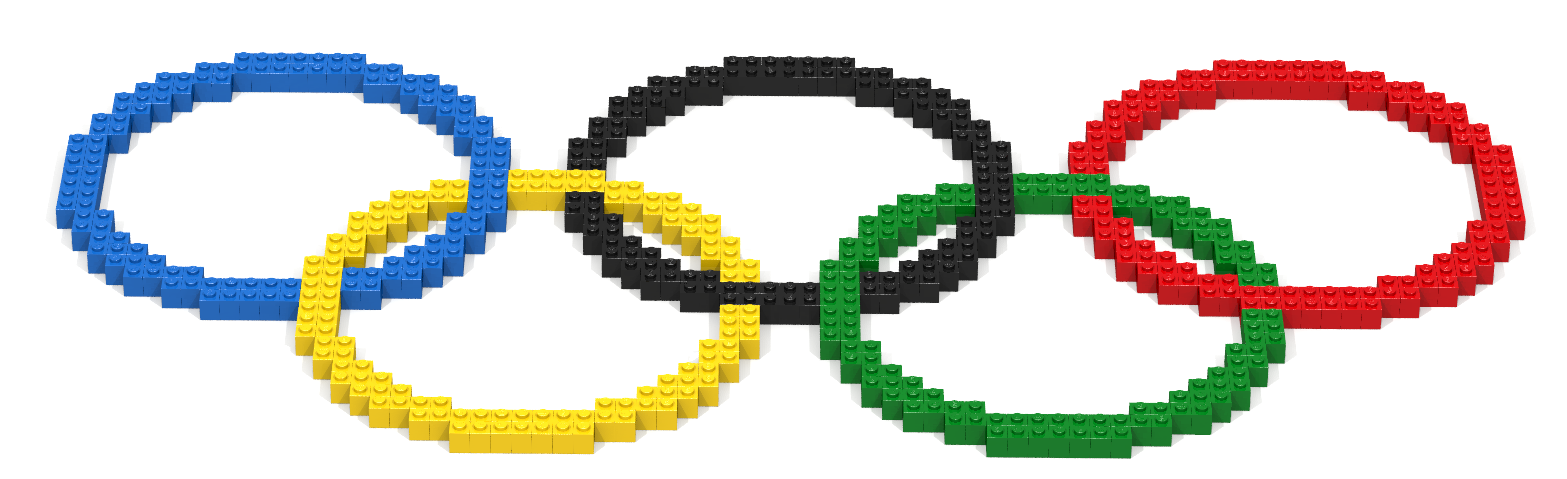}
\caption{The (interlocking) Olympic rings created using rings, clipping and setMySpace2D. }~\label{fig:olympic-rings}
\end{figure}

A brief overview of the Vitruvia concept sequence follows (for a more detailed discussion see \cite{2015Winter:Bricklayer} or visit \url{bricklayer.org/vitruvia}). Concept 1 establishes an understanding of two dimensional coordinate systems. Exercises involve the placement of $1 \times 1$ \lego\ bricks\footnote{Technically the dimension of such bricks is $1 \times 1 \times 1$.} (aka ``bit bricks'') in cells having specific coordinates. Concept 2 briefly touches on property-based construction of artifacts. For example, the even/odd property of coordinates can be used to place black/white bricks thereby creating a checkerboard pattern. Concepts 3--4 establish a natural language vocabulary for specifying the placement and orientation of \lego\ bricks having standard shapes. Concepts 5--6 establish a translation between natural language specifications and \bricklayer\ function calls. More specifically, at this stage artifacts are constructed through a sequence of calls to \bricklayer\ functions that ``put'' bricks having standard shapes and colors at specified coordinates. Concepts 7--9 focus on \emph{overwriting} (the result of brick collisions) as well as debugging basics. Concepts 10--12 introduce \emph{function declarations} and \emph{offsets}. Concept 13 introduces the \emph{ring} and the \emph{circle} as well as the idea of \emph{curried functions}. Concept 14 improves the ``put'' function by parameterizing it on shape, brick type, and coordinate. It is now possible for a single ``put'' function call to create a rectangle of arbitrary dimensions. Concept 15 introduces a function for creating ``smooth'' lines. Concept 16 discusses clipping and how it can be used to advantage within a program. Concept 17 introduces a function called \emph{setMySpace2D} which enables partitioning of \bricklayer's virtual space. This partitioning enables clipping to be used in more creative ways and also opens up the possibility of creating multi-user artifacts. Also, artifacts containing interlocking rings, like the Olympic symbol shown in Figure \ref{fig:olympic-rings}, can be created using \bricklayer's ring and myspace functions.

Concept 18 introduces the \emph{let-block} as a mechanism for structuring code. Concept 19 transitions the programmer into three dimensional space and introduces the three dimensional version of the ``put'' function. Concept 20 focuses on \emph{conditional expressions} and \emph{brick functions}. Concept 21 focuses on nested conditional expressions as well as the logical operator \emph{andalso}. Concept 22 focuses on nested conditional expressions as well as the logical operator \emph{orelse}. Concept 23 introduces the \emph{traverseWithin} iterator.

Vitruvia is extensible and we have plans for additional concepts as well as other enhancements such as the creation of leaderboards and intelligent testing.

\subsection{Minecraft}\label{section-minecraft}

The \bricklayer\ library is integrated with Minecraft through a CanaryMod plugin called RasberryJuice. This plugin is a PC implementation of the key functionality of the Minecraft: Pi Edition modding API (the Pi Edition is a free version of Minecraft developed by Mojang for the Rasberry Pi computer). The RasberryJuice plugin provides a function, called \emph{setBlock}, capable of placing a block at a particular position in a Minecraft world. In order for \emph{setBlock} to work, the Minecraft world in which the block is to be placed must be up and running on the CanaryMod server. Another important RasberryJuice function is called \emph{setPos} which enables a player to be placed at a particular location in a Minecraft world.

It is primarily through the RasberryJuice functions setBlock and setPos that \bricklayer\ interacts with a Minecraft world (residing on a CanaryMod server).

\begin{figure}[htb!]
\centering
\includegraphics[width=\textwidth]{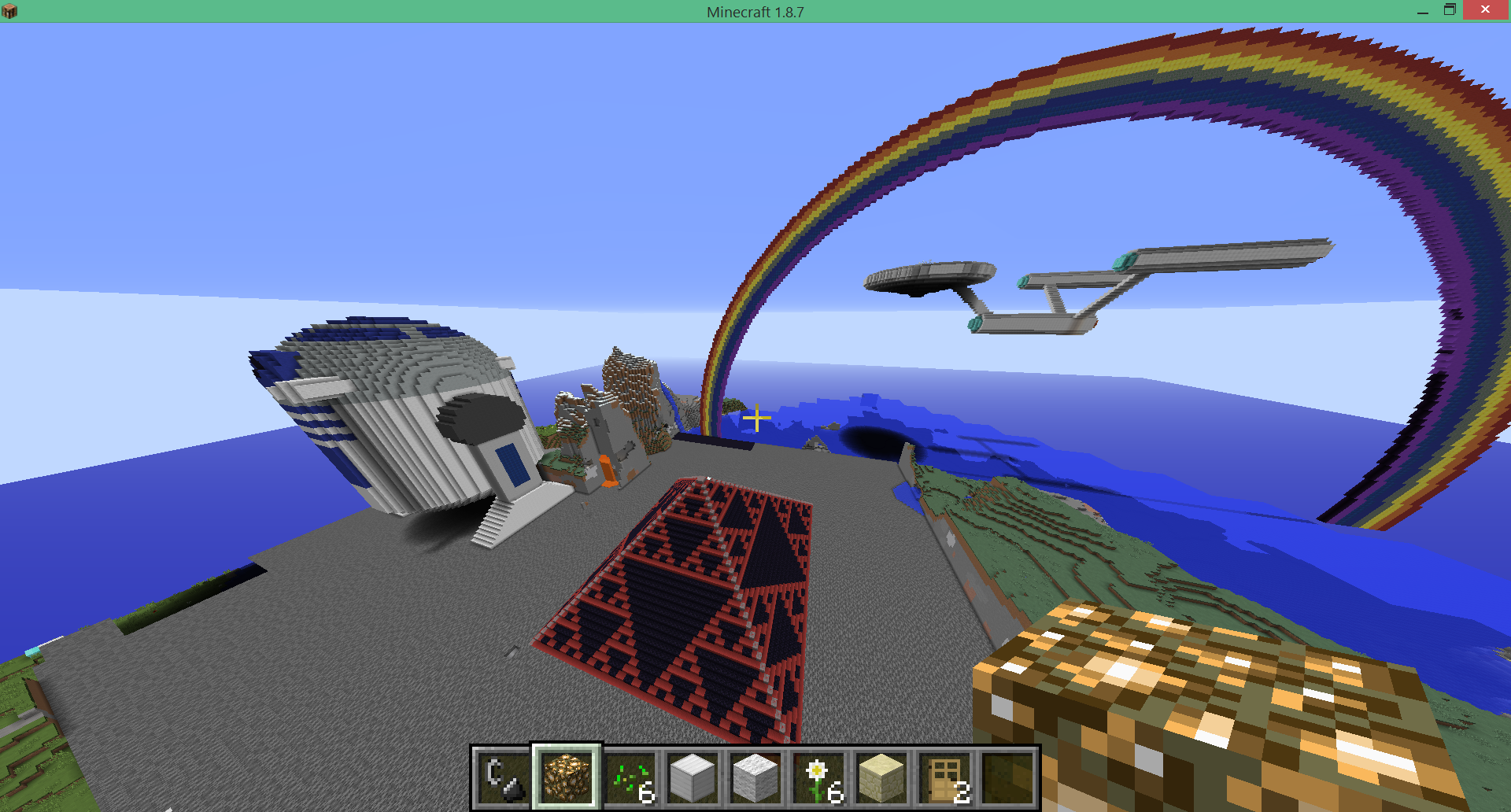}
\caption{\bricklayer\ artifacts in Minecraft: Worlds collide.}~\label{fig:minecraft}
\end{figure}

An artifact residing in Bricklayer's virtual space can be built in (i.e., exported to) a Minecraft world running on a CanaryMod server by simply changing the output function of a \bricklayer\ program from \emph{show} (which outputs the artifact to LDD) to \emph{showMC}. The showMC function will cause \bricklayer\ to create a Python program consisting of one setPos function call, to position your avatar (i.e., player), followed by a sequence of setBlock function calls, one function call for each cell in \bricklayer's virtual space occupied by a piece. This Python program is written to a file which is then automatically executed using a system-level command. The resulting artifact can then be seen using a Minecraft client -- provided it is connected to the CanaryMod server. From a conceptual standpoint, that is all there is to it. The rest is engineering. Figure \ref{fig:minecraft} shows a Minecraft world populated with several \bricklayer\ artifacts.

In addition to \emph{showMC}, the \bricklayer\ library provides several functions facilitating the construction of Minecraft artifacts. For example, functions are provided for (1) establishing correspondences between the coordinate system of the \bricklayer\ virtual space and the coordinate system of Minecraft, (2) specifying Minecraft-specific coordinate offsets, and (3) positioning \bricklayer\ artifacts at specific Minecraft coordinates.

It should be noted that artifacts viewed in LDD are ``stateless'' in the sense that if an artifact does not appear as desired one simply closes LDD, modifies the \bricklayer\ program, executes the new program, and views the modified artifact (i.e., the previously viewed artifact no longer exists). In contrast, artifacts viewed in Minecraft are ``stateful'' -- they remain in the Minecraft world until they are explicitly erased. To facilitate the artifact development cycle in Minecraft, the \bricklayer\ library provides a function for erasing \bricklayer\ artifacts from a Minecraft world.

Because of ``statefullness'' and the overhead (e.g., explicit user actions) associated with Minecraft client viewing, a feedback loop involving the design of an LDD artifact is much simpler (and faster) than the feedback loop involving the design of a Minecraft artifact. For this reason, it is highly beneficial to prototype artifacts in LDD before moving them to a Minecraft world. In this way, the concept of prototyping naturally finds its way into \bricklayer\ coding.

There is a problem though: LDD has difficulty displaying artifacts containing more than 25K pieces\footnote{In special cases, LDD is able to display artifacts having 60K pieces and sometimes even a little bit more. We hope that this limit will be increased in future versions of LDD.}. In contrast, \bricklayer\ artifacts can be constructed in Minecraft worlds containing up to 450K pieces. This difference naturally motivates the parameterization of an artifact on its size. The construction of smaller artifacts can be prototyped using LDD and the size of the artifact can be increased when it is moved to a Minecraft world.

Currently, the \bricklayer\ library supports 97 \lego\ bricks and 343 Minecraft blocks. Default mappings are provided between \lego\ bricks and Minecraft blocks. This allows artifacts to be fluidly displayed using either LDD or Minecraft with minimal program modification (i.e., simply toggle between the \emph{show} and \emph{showMC} output functions). The \bricklayer\ library also permits users to define custom mappings between \lego\ bricks and Minecraft blocks allowing LLD distinctions of Minecraft blocks in an artifact-specific fashion.

\subsection{3D Printing}\label{section-3d-printing}
\begin{figure}[htb!]
\centering
\begin{subfigure}[b]{0.5\textwidth}
\includegraphics[width=\textwidth]{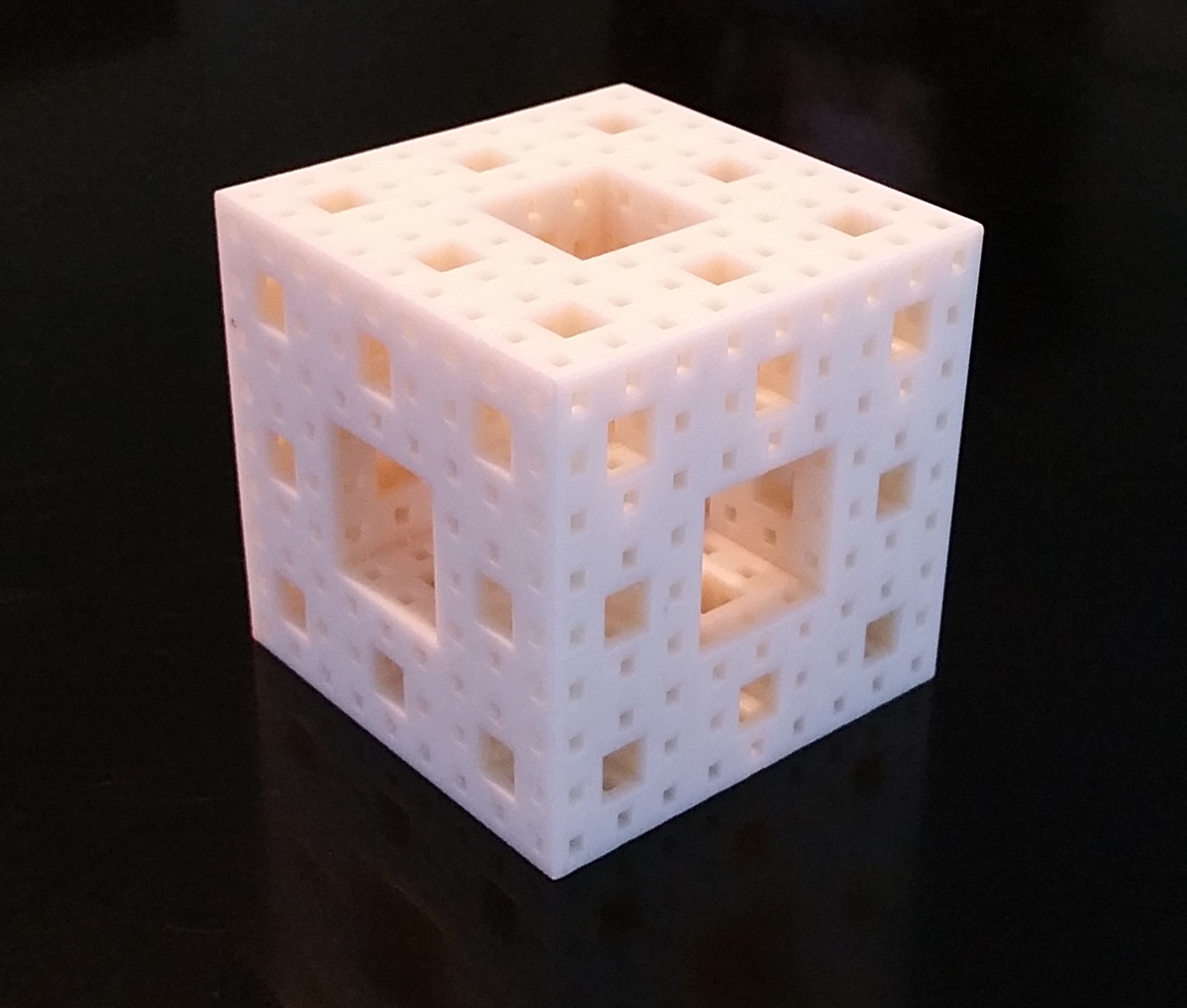}
\caption{A 3D printed Menger sponge consisting of 8000 bricks.}
\label{fig:3d-printed-menger}
\end{subfigure}
\qquad
\begin{subfigure}[b]{0.41\textwidth}
\includegraphics[width=\textwidth]{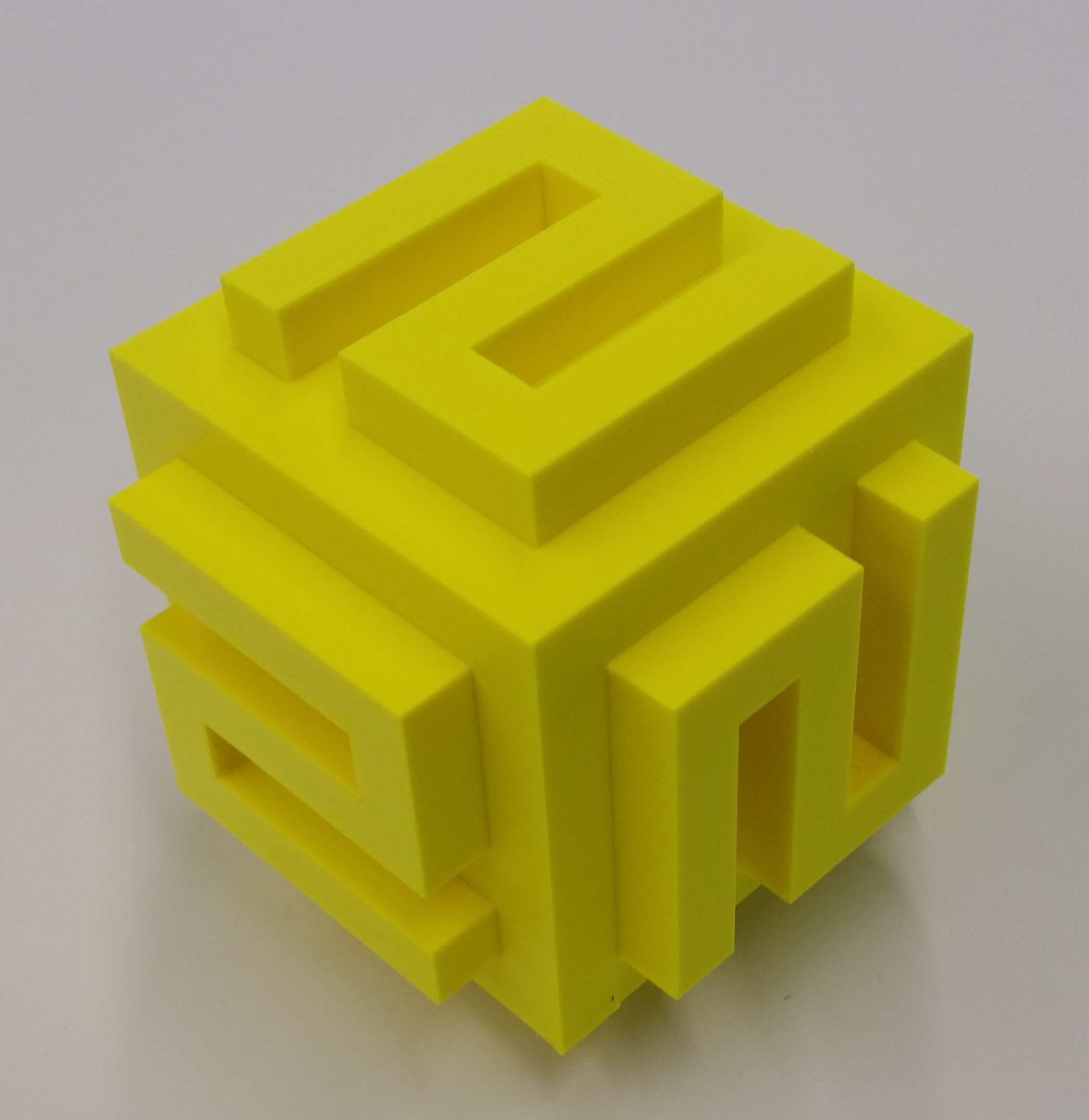}
\caption{A Wunderlich cube.}
\label{fig:wunderlich}
\bigskip
\end{subfigure}

\bigskip

\begin{subfigure}[b]{0.6\textwidth}
\includegraphics[width=\textwidth]{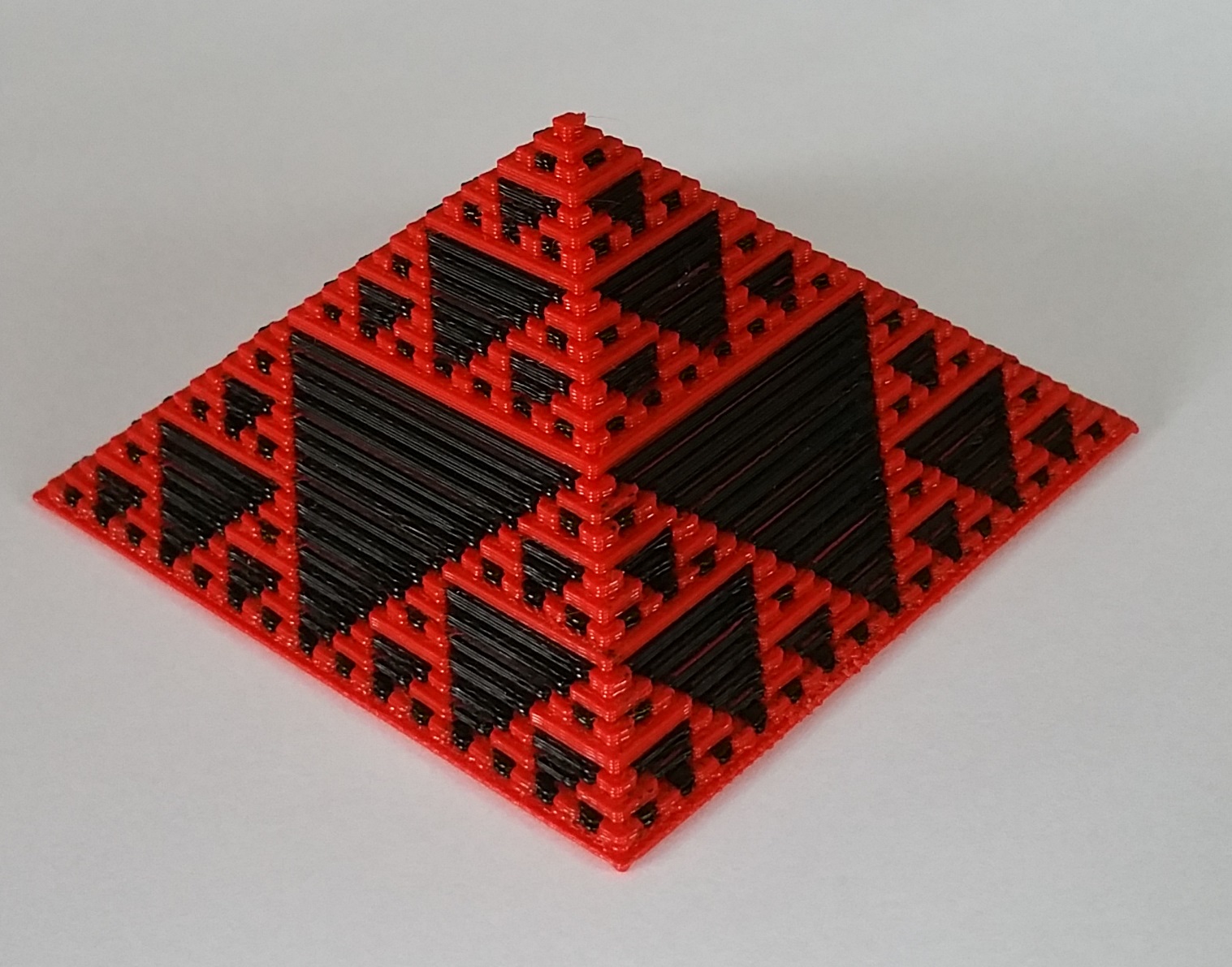}
\caption{A dual extrusion 3D print of the Sierpinski pyramid.}
\label{fig:wunderlich}
\end{subfigure}
\caption{\bricklayer\ 3D printed artifacts.}
\label{fig:3D-printed-artifacts}
\end{figure}

3D printing provides an educational dimension that can be highly engaging. Over the past 2 years the most frequent \bricklayer\ requests/suggestions have been related to 3D printing. Such requests have originated from researchers, teachers, students, and parents alike.

In response to these requests, the \bricklayer\ library has been extended to support the output of artifacts in a STereoLithography (STL) format. The artifacts described by such files can be viewed and 3D printed using viewers such as 3D Builder (Windows OS) and Cura. Figure \ref{fig:3D-printed-artifacts} shows a variety of 3D printed \bricklayer\ artifacts. It is worth mentioning that 3D printed \bricklayer\ artifacts are constructed from cubes (not \lego\ bricks).

\subsection{LDraw}\label{section-ldraw}

\bricklayer\ can output artifacts in the LDraw file format thereby enabling artifacts to be viewed using LDraw. The LDraw viewer has capabilities that distinguishes it from the LDD viewer. First, LDraw can display an artifact having 250K bricks -- in contrast to LDD which begins to have difficulty when artifacts contain more than 25K pieces. Second, in LDraw it is possible to continuously rotate an artifact around an arbitrary axis (e.g., the x-axis). Though useful, this behavior is not possible in LDD, which only permits unrestricted rotation around the y-axis. The 3D printed Wunderlich cube shown in Figure \ref{fig:wunderlich} was originally validated in LDD by rotating the cube around the y-axis. The artifact looked good and was 3D printed. Only later did we realize that the cube we had constructed contained an error -- a costly mistake -- which would have been revealed had we rotated the cube around the x-axis. A second (and correct) version of the Wunderlich cube was validated using the rotational capabilities of LDraw prior to 3D printing.

\subsection{Brickr}\label{section-brickr}

\begin{figure}[htb!]
\centering
\begin{subfigure}[b]{0.4\textwidth}
\includegraphics[width=\textwidth]{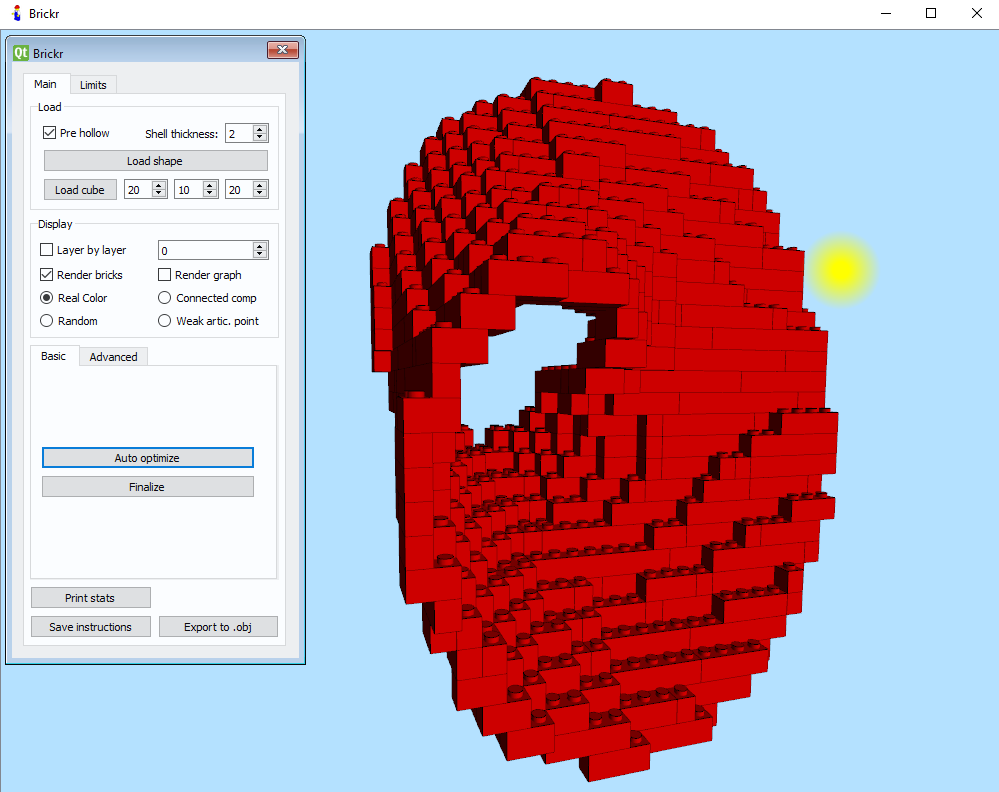}
\caption{A Brickr solution to the \lego\ construction problem for the Moebius strip.}~\label{fig:brickr}
\end{subfigure}
\qquad
\begin{subfigure}[b]{0.5\textwidth}
\includegraphics[width=\textwidth]{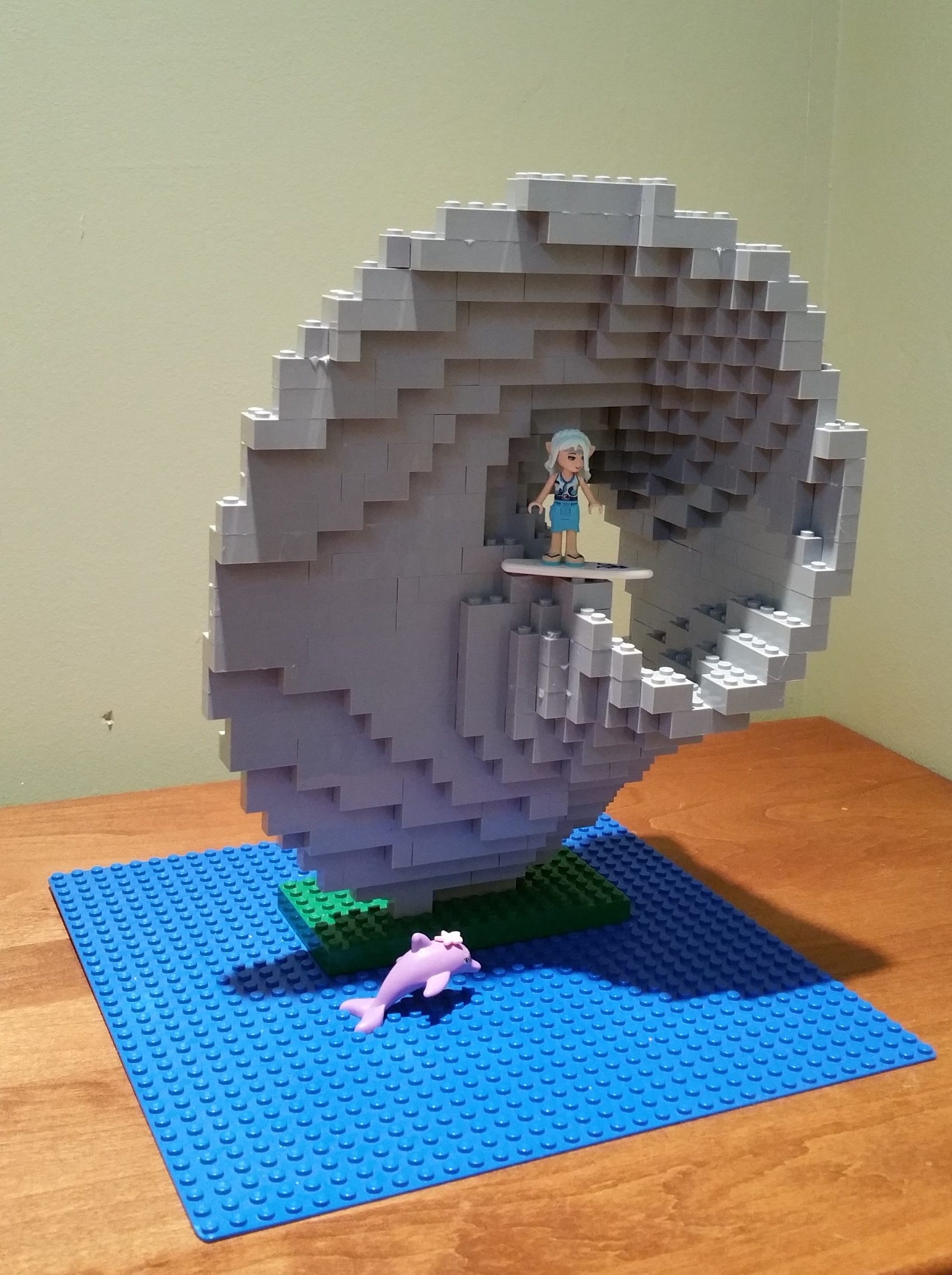}
\caption{A Moebius strip constructed from \lego\ bricks.}
\label{fig:lego-moebius}
\bigskip
\end{subfigure}
\caption{Using \bricklayer\ and Brickr to create a Moebius strip in physical space.}
\label{fig:moebius-problem}
\end{figure}

A well-known problem, called \emph{The \lego\ Construction Problem}, concerns itself with the construction of 3D objects using a given set of standard \lego\ bricks \cite{timcenko:LEGO-Construction-Problem}. The starting point for this problem is a model of a 3D object consisting entirely of $1 \times 1 \times 1$ \lego\ bricks. Such a model is referred to as a ``legoized'' representation of the 3D object. The goal is to construct a \lego\ artifact in physical space corresponding to the ``legoized'' model. There are two primary objectives the construction must satisfy: (1) the \lego\ artifact must have the same shape as the ``legoized'' model, and (2) the \lego\ artifact must hold together under the forces of gravity. Secondary objectives involve issues such as (1) optimization of brick shape usage, and (2) brick orientations to increase the strength of the \lego\ artifact.

At this time, all artifacts produced by \bricklayer\ are in ``legoized'' form. This makes them ideal inputs for tools, such as Brickr\cite{otaduy:brickr}, that focus on the \lego\ construction problem.

To support integration with Brickr, a special \bricklayer\ output function has been implemented, called \emph{showBinvox}. This function will output an artifact in a binvox format -- a file format suitable for input to Brickr. As a result, Brickr is integrated into the \bricklayer\ ecosystem.

In Brickr, the \lego\ construction problem can be solved by simply pressing the \emph{Auto optimize} button. Additionally, Brickr has the ability to produce step-by-step instructions showing how to create the resulting \lego\ artifact manually. This opens the door to a fascinating problem domain involving the construction of non-trivial mathematical objects including: the Moebius strip, the Klein bottle, the trefoil knot, and minimal surfaces such as the Enneper minimal surface. The feasibility of constructing such creations in \lego\ was demonstrated by Andrew Lipson \cite{lipson}\cite{lipson:wired}, a mathematician who earned a PhD in knot theory from Cambridge. Lipson creates C/C++ programs that construct ``legoized'' models of mathematical objects and then solves the \lego\ construction problem manually -- which he says is ``...where the fun is''. In contrast, we use \bricklayer\ to create a ``legoized'' model and Brickr to solve the \lego\ construction problem. Those who want to share in Lipson's fun can output Brickr instructions for ``legoized'' artifacts without first pressing Brickr's \emph{Auto optimize} button.

Figure \ref{fig:brickr} shows a Brickr solution to the \lego\ construction problem for a ``legoization'' of a Moebius strip created by a \bricklayer\ program. Brickr instructions for how to manually create this artifact in \lego\ were produced and the authors then (of course) ordered the requisite parts from \lego\ and built the \lego\ artifact. The result is shown in Figure \ref{fig:lego-moebius}.

\subsection{Bricklayer-Lite}\label{section-bricklayer-lite}


\begin{figure}[htb!]
\centering
\begin{subfigure}[b]{\textwidth}
\includegraphics[trim = 0mm 0mm 0mm 0mm, clip, scale=0.40]{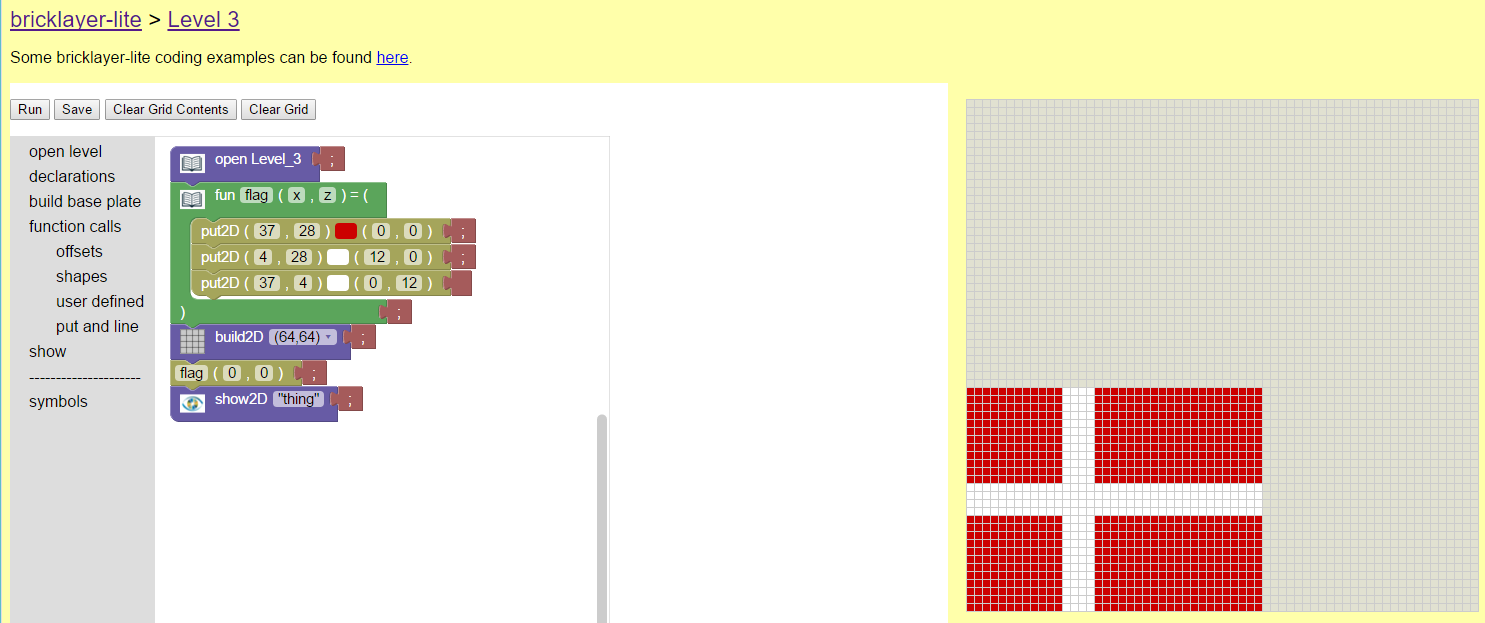}
\caption{An example of a bricklayer-lite program and the artifact it creates.}
\label{fig:bricklayer-lite-artifact}
\end{subfigure}

\bigskip

\begin{subfigure}[b]{\textwidth}
\begin{lstlisting}
(* generated by Bricklayer Lite version 0.9 *)

open Level_3;

fun flag (x,z) =
	(
		put2D (37,28) RED (0,0);
		put2D (4,28) WHITE (12,0);
		put2D (37,4) WHITE (0,12)
	);

build2D (64,64);

flag(0,0);

show2D "flag";
\end{lstlisting}
\caption{Auto-generated Bricklayer code produced when running a bricklayer-lite program.}
\label{fig:bricklayer-lite-autogen}
\end{subfigure}
\caption{Coding using bricklayer-lite.}
\label{fig-bricklayer-lite}
\end{figure}

Bricklayer-lite is a web app that provides a visual block-based coding environment for creating \bricklayer\ programs. Bricklayer-lite is implemented
using the Blockly library developed by Google. Currently, the functions available in Levels 1--3 have been implemented in bricklayer-lite. This was accomplished by defining custom Blockly blocks and then defining their semantics in JavaScript. The semantics includes a state machine which tracks various properties relating to well-formed programs.

When executing a well-formed bricklayer-lite program, the JavaScript semantics include a function that generates executable \bricklayer\ code corresponding to the bricklayer-lite program. The \bricklayer\ program is displayed in a web form below the bricklayer-lite program. An example is shown in Figure \ref{fig-bricklayer-lite}. Figure \ref{fig:bricklayer-lite-artifact} shows a bricklayer-lite program and the artifact it creates. Figure \ref{fig:bricklayer-lite-autogen} shows the Bricklayer program, auto-generated by bricklayer-lite, corresponding to the program shown in Figure \ref{fig:bricklayer-lite-artifact}.

Bricklayer-lite runs on any web browser and can be accessed from any device such as a smart phone, tablet (e.g., iPad), or laptop. Bricklayer-lite programs are pictorial in nature and are constructed by fitting together blocks. Depending on the technology used (e.g., laptop, smart phone), blocks can be selected via mouse-clicks or by touch. Some, though very little, typing (e.g., data entry) is required when constructing a bricklayer-lite program. For example, to place a brick at a specific location requires entering, via a keyboard, the coordinates of that location.

Bricklayer-lite programs can be executed by selecting the run button. They can be saved to the cloud and retrieved at a later time as well as from other devices.

\subsection{The Lace Animator}\label{section-animator}

The \bricklayer\ ecosystem contains a variety of (pre-coding) un-plugged exercises which focus on the understanding and discovery of algorithms for creating 2D patterns on graph paper. The simplest class of pattern involves the repetition of a set of seed patterns whose size remains constant. We refer to such patterns as \emph{arithmetic patterns}. Exercises involving arithmetic patterns include tessellations of the a 2D plane using pentomino shapes. It should be noted that the spacing of shapes in regular patterns follow arithmetic progressions.

\begin{figure}[htb!]
\centering
\includegraphics[width=0.7\textwidth]{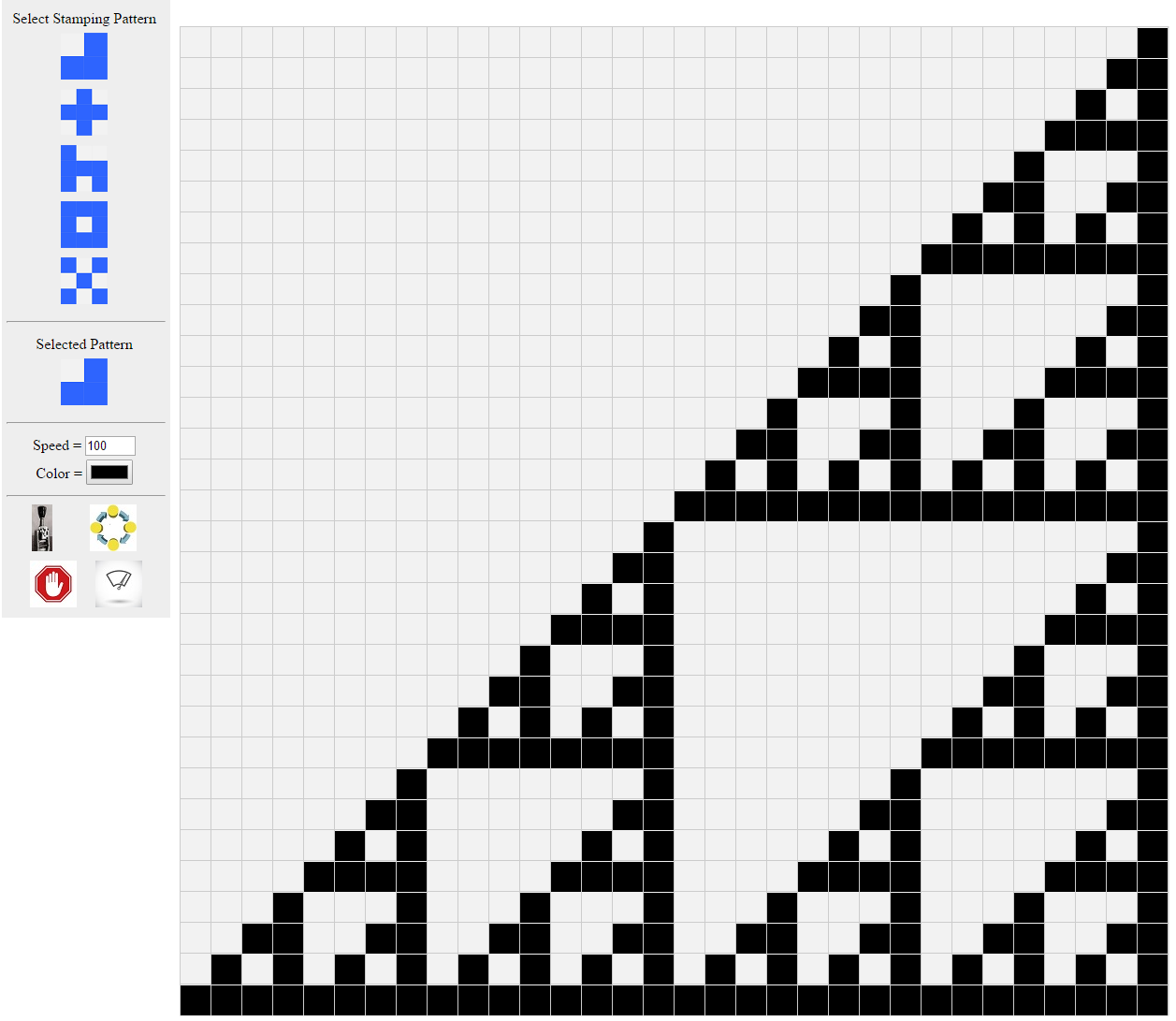}
\caption{The result of a lace pattern animation.}~\label{fig:animator}
\end{figure}

The next class of pattern we call \emph{geometric patterns}. These are patterns whose spacing follow geometric progressions. As a result, patterns get larger/smaller as they repeat.  Fractals fall into this category of pattern. We have coined the term \emph{lace} to refer to a class of geometric patterns created from a seed shape and a stamping pattern. A \emph{lace} is a structure that is connected but not a line.

Empirical evidence suggests that people of all ages have an initial difficulty constructing laces -- they have a tendency to employ an arithmetic approach (rather than a geometric approach) to constructing lace patterns. In particular, students want to repeat one pattern over and over without changing its size. To help students understand the nature of the algorithm that underlies lace patterns we have developed an interactive web app called the \emph{Lace Animator} which shows how a variety of laces can be constructed using linear steps (one seed shape at a time) as well as geometric steps (one pattern at a time). Figure \ref{fig:animator} shows a lace created using the Lace Animator.

\section{Outreach}\label{section-education}

\bricklayer\ has been used for a variety of K-12 outreach efforts including those listed below.

\begin{itemize}
\item Summer Workshops -- in the summer of 2014 and again in 2015 \bricklayer\ was used to teach coding to elementary and middle school students. The educational format was a 3 hour/day 5 day workshop. The venue was the Techademy, a summer tech program offered by the College of Information Science \& Technology at the University of Nebraska at Omaha.

\item Coding Club -- An after school \bricklayer\ coding club was held at Rockbrook Elementary School from November, 2014 -- May, 2015. The club met once a week for an hour and culminated with a parent's ``Art and Pizza Night'' event in which students displayed on the school walls the artifacts they created alongside their code.

\item Exhibits
\begin{itemize}
\item NE SciFest -- The Nebraska Science Festival is a major annual event in Nebraska hosted by the University of Nebraska Medical Center. Its purpose is to stimulate interest in science. In 2015, \bricklayer\ had an exhibit at the NE SciFest -- a 2 day event which was held that the Durham Museum in Omaha. It was estimated that approx 3000 students (mostly from elementary school) were in attendance.
\item Lauritzen Gardens -- a botanical center in Omaha periodically hosts a multi-week \lego\ Artists Expo where artists present life-size \lego\ sculptures. In 2015, during this expo, \bricklayer\ was invited to have a 1-day exhibit on the \lego\ robotics day.
\item Omaha Children's Museum -- Beginning in May 2015, the Omaha Children's Museum had a multi-month \lego\ exhibit. \bricklayer\ was invited to hold an exhibit on the opening day of the Museum's \lego\ exhibit.
\end{itemize}
\item CoderDojo -- In the summer of 2015, \bricklayer\ was used in a topics in mathematical computing course for in-service secondary school math teachers. The course included a 4-week CoderDojo session in which the secondary school math teachers taught \bricklayer\ to a group of students ranging in age from 7 to 17.
\item Gifted Education -- In the spring of 2016, a 10-week pilot study was undertaken in which six facilitators taught a \bricklayer-based curriculum to approximately 120 $4^{th}$ grade students enrolled in the Gifted and Talented Program of the Omaha Public School system.
\end{itemize}

\section{Conclusion}\label{section-conclusion}

The \bricklayer\ ecosystem has a demonstrated appeal to students in the K-12 spectrum. It continues to evolve in order to more effectively engage all who wish to learn to code or sharpen their coding abilities. \bricklayer\ code has strong mathematical underpinnings including coordinate systems,  geometric objects, arithmetic and geometric progressions that govern the structure of tessellations and fractals, and Boolean algebra. \bricklayer\ artifacts are artistic in nature. This engages visual thinking and spatial reasoning -- cognitive processes which are closely correlated with STEM abilities. Last but not least, through \lego\ and 3D printing, \bricklayer\ spans the physical and virtual worlds.

\bibliographystyle{eptcsini}
\bibliography{bricklayer_ecosystem}

\newpage

\end{document}